\documentclass[aps,prm,twocolumn,groupedaddress,amsmath,amssymb]{revtex4-2}

\usepackage{graphicx}
\usepackage[colorlinks=true, citecolor=blue, urlcolor=blue]{hyperref}
\usepackage{color}
\usepackage{url}

\begin{document}

\title{First-principles study of the phase competition, mechanical and piezoelectric properties of pseudo-binary (SiC)$_{1-x}$(AlN)$_{x}$ alloy}

\author{Laszlo Wolf}                                                                                                     
\affiliation{Colorado School of Mines, Golden, CO 80401, USA}

\author{Goeff L. Brennecka}
\affiliation{Colorado School of Mines, Golden, CO 80401, USA}

\author{Vladan Stevanovi\'{c}}
\email{vstevano@mines.edu}                                                                     
\affiliation{Colorado School of Mines, Golden, CO 80401, USA}

\date{\today}

%
\begin{abstract}
The ongoing search for new piezoelectric materials offering adequate balance between piezoelectric response and other application-relevant properties has lead to the investigation of various alloy systems. In this work we study the alloy of the widely used AlN with SiC for their relative abundance, current use in other electronics applications and expected phase competition between wurtzite and other polymorphs, the kind of which has lead to some of the most interesting results notably between AlN and ScN.  Here the pseudo-binary (SiC)$_{1-x}$(AlN)$_{x}$ alloy is studied from first-principles over the entire composition range. Relevant crystalline phases are identified using the First-Principles Random Structure Sampling approach which, in accordance with previous bulk experiments, finds wurtzite, zincblende and rhombohedral phases to be the only statistically relevant phases of the alloy. Further study of these phases is done through Special Quasi-random Structures (SQS) and, in the case of the wurtzite phase, predictions of the stiffness, piezoelectric and dielectric tensors. Analysis of these tensors is done through the scope of a Bulk Acoustic Wave (BAW) filter application, where trends and trade-offs between the c-axis acoustic velocity and piezoelectric response enable identification of relevant compositions. 
\end{abstract}
%
\maketitle
%

\section{Introduction}\label{sec:introduction}
%
Alloying has been a common procedure for improving piezoelectric materials. Notable examples are the AlN-based alloys~\cite{starttAEM-2023}. The aim has mostly been toward increasing the relatively low piezoelectric constant of the wurtzite AlN phase ($d_{33}\approx5$ pC/N) achieved through alloying with scandium~\cite{k2_def,ambacher_JAP_2023}. However the scarcity of Sc has greatly increased the demand for alternatives, preferably more affordable and easily implemented with current fabrication techniques such as Complementary Metal-Oxide Semiconductors (CMOS) for example.

In this work the choice of exploring the alloy between AlN and SiC is motivated by several factors. First, this alloy has already been synthesized both as a bulk material and a thin film~\cite{alloy_1,alloy_2,alloy_3,alloy_4,alloy_5,alloy_6}. SiC is also readily available, relatively inexpensive, already present in micro-electro-mechnical systems where its used for its resistance to harsh conditions specifically at high temperature. It is also known for its polymorphism and the phase competition between different crystalline structures. Phase competition in AlN-based alloys has been the synonym for surprising piezoelectric property evolution and is thought to be at the origin of the surprising results of the AlScN alloy~\cite{tasnadiPRL_2010}. While the piezoelectric response usually takes the focus of materials development, other properties such as the speed of sound is of high importance in application such as resonators~\cite{BAW_1,BAW_2}. The high acoustic speed of SiC makes it attractive in that regard.

This work presents a study of the pseudo-binary alloy (SiC)$_{1-x}$(AlN)$_{x}$ through several steps in the following order: (i) identification of relevant phases and phase competition analysis, (ii) mechanical property calculations for the most relevant phase, (iii) comparison with experimental data, and finally, (iv) the analysis of the calculated electro-mechanical properties. To identify the statistically relevant phases of the alloy we perform First-Principles Random Structure Sampling (FPRSS), which consists of the generation of a large sample of random structures and their subsequent first-principles relaxations to the cloasest local minima on the Potential Energy Surface (PES) of the system. It has been shown ~\cite{jonesPRB-2017,stevanovicPRL-2016} that the frequencies of occurrence of various structure-types in the random samplings can be used to predict likelihood for experimental realizations of identified phases. The FPRSS was successfully used to reproduce known polymorphism in crystalline systems~\cite{stevanovicPRL-2016, jonesPRB-2017,jankouskyPRM-2023}, create structural models for amorphous phases~\cite{jonesNPJ-2020,wolfJAP-2025} and to elucidate polymorphism-linked phenomena in thin-films~\cite{woodsrobinsonPRM,zakutayevNS-2024}. 

Using FPRSS in this work for three different alloy compositions $(x=0.25, 0.5, 0.75)$ we have identified wurtzite, zincblende and rhombohedral as the 3 both energetically and statistically relevant phases, more likely than any other to be experimentally realized. The Special Quasi-random Structures (SQS)~\cite{zungerPRL}  used to approximate the random alloy provided the bases for the subsequent phase competition analysis, results of which reveal the wurtzite phase as the thermodynamically most stable phase in accordance with experimental results. Finally, the stiffness, piezoelectric and dielectric tensors are evaluated for the wrutzite phase across the entire composition range. From these tensors we obtain the chemical trends in the c-axis speed of sound and piezoelectric response as well as in the electromechanical coupling coefficient. These properties are of specific interest for BAW filters and allow identification of the compositions having the optimal combination of the piezoelectric properties and the speed of sound. 
%
\section{Methods}\label{sec:methods}
%

\subsection{First-principles random structure sampling}\label{ssec:methods_FPRSS}
%
The creation of the random structures of a given size (number of atoms) and a given chemical composition is done through following 3 steps: (i) a random cell is created with random lattice parameters ($a,b,c,\alpha,\beta,\gamma$), (ii) atoms are distributed within the cell with an algorithm designed for homogeneity of the distribution and also favoring cation-anion coordination in relevant cases, and (iii) first-principles relaxation of the structures to the closest local minimum~\cite{stevanovicPRB_2012}.

 In the relaxations, a standard first-principles setup is employed. The electron-electron interactions are treated using the PBE~\cite{perdewPRL} exchange-correlation functional and the projector augmented wave (PAW)~\cite{blochPRB} method as implemented in the VASP computer code~\cite{kresse_PRB_1999}. Automatic generation of the $\Gamma$-centered ${\bf k}$-point grid is employed with the $R_k$ value of 20. 
 All degrees of freedom, including volume, cell shape and atomic positions, are relaxed using the conjugate gradient algorithm~\cite{teter_PRB_1989}. Volume and cell-shape relaxations are restarted at least four times for the real-space grid to be re-created. Structural relaxations are considered converged when maximal force on any atom gets below 0.02 eV/{\AA}, total energy stops changing by more than $10^{-6}$ eV and the hydrostatic pressure drops below 0.5 kbar. 
 
 Using this procedure, the potential energy surface of the system is probed at three different compositions ($x=0.25,0.5,0.75$) where a total of ~5,000 random structures is generated and relaxed at each composition. The structures used here contain 24 atoms per previous FPRSS size dependance studies\cite{jonesNPJ-2020,zakutayevNS-2024,wolfJAP-2025}. To obtain accurate statistics (frequencies of occurrence) one must sort relaxed random structures into groups with the same structure-type. Because of the possible Al-Si disorder present in different structures the grouping  is done based on the symmetry (the space group number) of the underlying structure and the local first-shell coordination. The symmetry of the underlying (parent) crystal structure is obtained by labeling all cations with the same label and using the spglib library~\cite{spglib} with a tolerance of 0.6 {\AA}. The first-shell coordination is obtained using the same spatial tolerance and a count tolerance of 0.1. For the higher occurring groups further confirmation of the symmetry of the structures was done by analyses of simulated powder XRD spectrum as well as a visual one.
 
\subsection{Special Quasi-random Structures (SQS)}\label{ssec:methods_SQS}
%
\begin{figure}
\includegraphics[width=\linewidth]{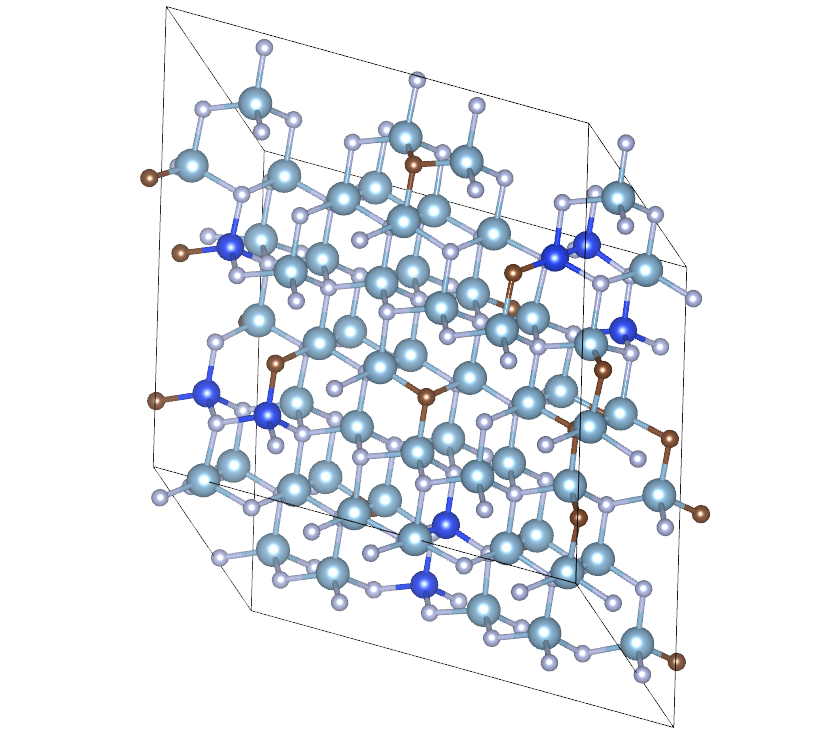}
\caption{\label{fig:strc} An example of a 128 atom (SiC)$_{0.125}$(AlN)$_{0.875}$ SQS relaxed using the first-principles methods described in Section~\ref{sec:methods}. Light blue, blue, white and brown spheres represent Al, Si, N and C atoms, respectively. }
\end{figure}

In order to approximate the occupation disorder expected in random alloys,  atom distribution over the cation and anion sublattices are obtained using the Monte-Carlo Special Quasi-random Structures (MC-SQS) program as implemented in the ATAT package~\cite{vandewalleCALPHAD}. These structures were created with 128 atoms for the wurtzite or rocksalt phases and 144 atoms for the rombohedral phase (see Fig.\ref{fig:strc}). A step in composition of $\Delta x = 0.125$ was used for each phase to sample the whole composition range of the alloy. For the Monte-Carlo procedure, clusters were considered up to the 5th neighbors for the pair interactions and up to the 3rd neighbor for the three body interactions. 

The procedure is launched 4 times for each instance with only the best 3 structures kept. These 3 structures are then relaxed through a similar DFT procedure as for the FPRSS ones with now a smaller total energy convergence criteria of $10^{-10}$ eV in order to facilitate the upcoming mechanical properties calculations.

Initial lattice parameters were determined from linear interpolation between previously relaxed pure compounds. One should also consider that even if the SQSs are created with an initial phase and symmetry, nothing guarantees that these will be conserved throughout the relaxation procedure. In that regard the phase and symmetry of the SQSs were identified again after relaxation following a similar procedure as with FPRSS stuctures.

\subsection{Mechanical Properties Calculations}\label{ssec:methods_mecha_props}
%
Relevant mechanical properties are calculated following VASP's recommendations such that the stiffness tensor is obtained through strain finite differences with the default step of $0.015$ \AA. The piezoelectric tensor is calculated from Density Functional Perturbation Theory (DFPT). 

 The isotropic Young's modulus and the isotropic Poisson's ratio are used for comparison with experimental measurements. These isotropic properties are obtained from the stiffness tensor through the Voigt and Reuss averages (high and low limits) of the bulk modulus $B_{V}$ and $B_{R}$ and the shear modulus $G_{V}$ and $G_{R}$~\cite{hill1952,dejong_SD_2015} with:
\begin{align}
	9B_{V} = ~&C_{11} +  C_{22}  +  C_{33}  + 2\left(  C_{12}  +  C_{23}  +  C_{31}  \right),\\
	\frac{1}{B_{R}} = ~&S_{11}  +  S_{22}  +  S_{33} + 2\left(  S_{12}  +  S_{23}  +  S_{31}  \right),\\
	15G_{V} =  ~&C_{11} +  C_{22} +  C_{33} -  C_{12} - C_{23} - C_{31} + \nonumber \\
	&+ 3\left(C_{44} + C_{55} + C_{66}\right), \\
	\frac{15}{G_{R}} = ~&4\left(S_{11} + S_{22} + S_{33} - S_{12} - S_{23} - S_{31}\right) + \nonumber\\
	&+3\left( S_{44} + S_{55} + S_{66}\right),
\end{align}
where the $S_{\alpha\beta}$ are components of the compliance tensor $\bf{S}=\bf{C}^{-1}$. In the above equations the stiffness tensor is represented by a $6\times6$ matrix through the Voigt notation such that $xx\rightarrow1$, $yy\rightarrow2$, $zz\rightarrow3$, $yz\rightarrow4$, $xz\rightarrow5$, $xy\rightarrow6$. This way the stiffness tensor $\bf{C}$ has components $C_{\alpha\beta}(\alpha,\beta = 1,\dots,6)$. The isotropic Young's modulus $E$ and the isotropic Poisson's ratio $\nu$ are then obtained from:
\begin{align}\label{eq:E_nu}
	\nu &= \frac{3B_{VRH}-2G_{VRH}}{6B_{VRH}+2G_{VRH}} & E &= 3B_{VRH}\left(1-2\nu\right),
\end{align}
where $B_{VRH}=\frac{B_{V}+B_{R}}{2}$ and $G_{VRH}=\frac{G_{V}+G_{R}}{2}$ are averages between the high and low bounds named the Voigt-Reuss-Hill averages.

The speed of sound along any direction defined by the unit vector ${\bf n} = (n_1,n_2,n_3)$ is obtained by solving the Christoffel eigenvalue equation~\cite{jaeken2016},
\begin{equation}\label{eq:sound_speed}
	\sum_{ij} \left[ \left(\sum_{kl} n_{k}C_{iklj}n_{l}\right) - \rho v^{2} \delta_{ij} \right] s_{j} = 0
\end{equation}
where $i,j,k,l$ indices run over the three cartesian directions, $C_{iklj}$ is an 4-index element of the stiffness tensor, $v=|{\bf v}|$ is the norm of the phase velocity of the monochromatic plane wave polarized along ${\bf{s}} = (s_x,s_y,s_z)$ direction. Solving this eigenvalue equation offers three solutions two of them corresponding to tranversal modes while the third corresponds to the longitudinal one; the eigenvectors corresponding to the polarization direction and the eigenvalue to the associated velocities.

The piezoelectric stress tensor $\bf{e}$ is a $3\times 6$ matrix following the same Voigt notation as for the stiffness tensor.
\begin{equation}
\bf{e} \, = \,
\begin{pmatrix}
e_{11} & e_{12} & e_{13} & e_{14} & e_{15} & e_{16} \\
e_{21} & e_{22} & e_{23} & e_{24} & e_{25} & e_{26} \\
e_{31} & e_{31} & e_{33} & e_{34} & e_{35} & e_{36} \\
\end{pmatrix}.
\end{equation}
While calculations performed here return the piezoelectric stress tensor $\bf{e}$, the experimentally reported piezoelectric strain tensor $\bf{d}$ can be obtained through the compliance tensor $\bf{S}$~\cite{nye1986}.
\begin{equation}\label{eq:d33}
	d_{\alpha\beta} \, = \, \sum_{\gamma=1}^{6} e_{\alpha \gamma} S_{\gamma\beta}.	
\end{equation}

%
\section{Results}\label{sec:results}
%
%
\begin{figure*}
\includegraphics[width=\linewidth]{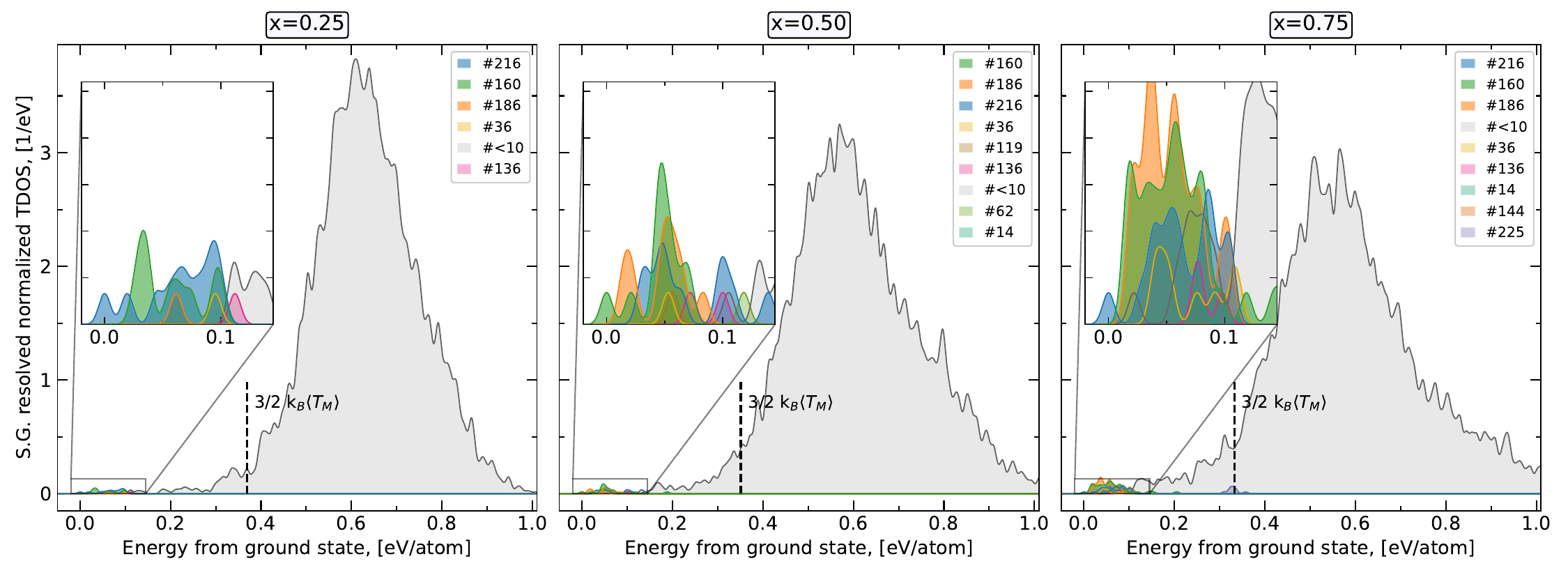}
\caption{\label{fig:RSL} Space group resolved thermodynamic density of states of the (SiC)$_{1-x}$(AlN)$_{x}$ alloy for $x=0.25$, $0.50$ and $0.75$. Each plot is derived from $\sim5000$ structures obtained from random structure sampling. Groups of structures with the same underlying/parent structure (see Section~\ref{ssec:methods_FPRSS} for details) are designated by their parent-structure space group numbers and are ordered in ascending (minimal) total energy in each legend. For clarity low symmetry structures (space group \#$<10$) have been grouped together. For better visualization, a Gaussian broadening of 0.005 eV has been applied to the data.}
\end{figure*}
%

\subsection{FPRSS identification of relevant phases}\label{ssec:results_RSL}
%
Identification of relevant phases of the (SiC)$_{1-x}$(AlN)$_{x}$ alloy was done by performing FPRSS at three specific compositions $x=0.25, 0.50, 0.75$. The resulting space group resolved Thermodynamic Density Of States (TDOS) are presented in Fig.~\ref{fig:RSL}. Like structures are grouped together and referred to by the space group number of the underlying/parent structure as described in Section~\ref{ssec:methods_FPRSS}. 

When analyzing the TDOSs in the three different compositions, one can observe two distinct regions in all of them. A low energy region, from $0.0$ eV to $\sim$0.3 eV/atom, where the high symmetry structures dominate, and a high energy region, centerred around $\sim 0.5-0.6$ eV/atom, dominated by the low symmetry structures. One can associate the high symmetry structures with different crystalline phases of the two parent compounds, while the low symmetry structures in aggregate can be thought of as representing the amorphous phase of the system \cite{jonesNPJ-2020,wolfJAP-2025}. 

The noticeable energy difference between the high symmetry structures and the amorphous phase is an indication of the system's preference to crystallize rather than grow amorphous. This conclusion is supported by previous studies of both crystalline phases~\cite{zakutayevNS-2024} and amorphous systems~\cite{wolfJAP-2025,Pshyk_AM:2025}  where the presence of relatively large energy separation between low-energy, high-symmetry (ordered) phases and low-symmetry states (s.g.~\# $< 10$) has been associated with a system favoring high symmetry crystalline phases. This can be further supported by estimating the thermal energy per atom at the approximate melting point of the alloy, $T_M(x) = (1-x)*T_{M}(SiC) + x*T_{M}(AlN)$ denoted in Fig.~\ref{fig:RSL} by the vertical dashed lines. Considering the system under some processing conditions at the temperatures close to the approximate melting point, structures far above dashed lines (the majority of the low-symmetry ones) would not be easily accessible to the system. This is in contrast to the high-symmetry structures which are located below the dashed lines thus indicating the tendency of the system to end up in one of those and for a crystalline phase.  Looking in more detail at the low energy region of each plot in Fig.~\ref{fig:RSL} the lowest energy structure groups are consistently those with space group numbers \#186, \#216 and \#160, that is, those with the underlying wurtzite, zincblende, and rhombohedral symmetry, respectively.  

In addition to their low energies, as shown in Table~\ref{table:RSL}, these three groups are always the most represented (the most frequently occurring) in the FPRSS, except for the \#186 structures at 25 \% AlN. The combination of these two arguments leads to the conclusion that experimental growth of the three different compositions studied here will most likely crystallize in one of the three phases identified (Hexagonal \#186, Cubic \#216, Rhombohedral \#160). This conclusion is in accordance with previous experimental results in growing both bulk and films of (SiC)$_{1-x}$(AlN)$_{x}$ alloys~\cite{alloy_1,alloy_2,alloy_6}. 

\begin{table}
\caption{\label{table:RSL} Frequency of occurence of high symmetry structures in the random structure sampling over different \%AlN stoichiometry. High-symmetry structure groups not present in this table appear significantly less than the 3 groups presented here.}
\begin{ruledtabular}
\begin{tabular}{cccc}
 &25 \%AlN & 50 \%AlN & 75 \%AlN \\
 \hline
\#186 (Wurtzite) & 1 & 10 & 35 \\
\#216 (Zinc-blende) & 11 & 10 & 20 \\
 \#160 (Rhombo.) & 9 & 11 & 33 \\
\end{tabular}
\end{ruledtabular}
\end{table}
%



\subsection{Mixing thermodynamics from SQS}\label{ssec:SQS}

Once the relevant phases are identified the special quasi-random structures (SQS) are generated for all three phases over a spectrum of compositions. Using SQS to model properties of a random alloy is more practical than the averaging over the structures obtained by the FPRSS. While the two methods are shown to produce very similar results \cite{novickPhysRevMat}, converging ensemble averages is much more computationally demanding in comparison to the phase identification.  Hence, in this work the FPRSS is used to identify the relevant phases and the tendency for crystallization, whereas the SQSs are employed to evaluate relevant properties of a random alloy.

Large, supercells reproducing the occupation disorder of the (SiC)$_{1-x}$(AlN)$_{x}$ alloys were created through the SQS procedure as detailed in Section \ref{ssec:methods_SQS}. Results shown in this section are averages over the three SQSs obtained at each instance. Unless indicated otherwise, the deviation between the three SQSs was found to be inconsequential, showcasing a good overall approximation of a random alloy. 

\begin{figure}
\includegraphics[width=\linewidth]{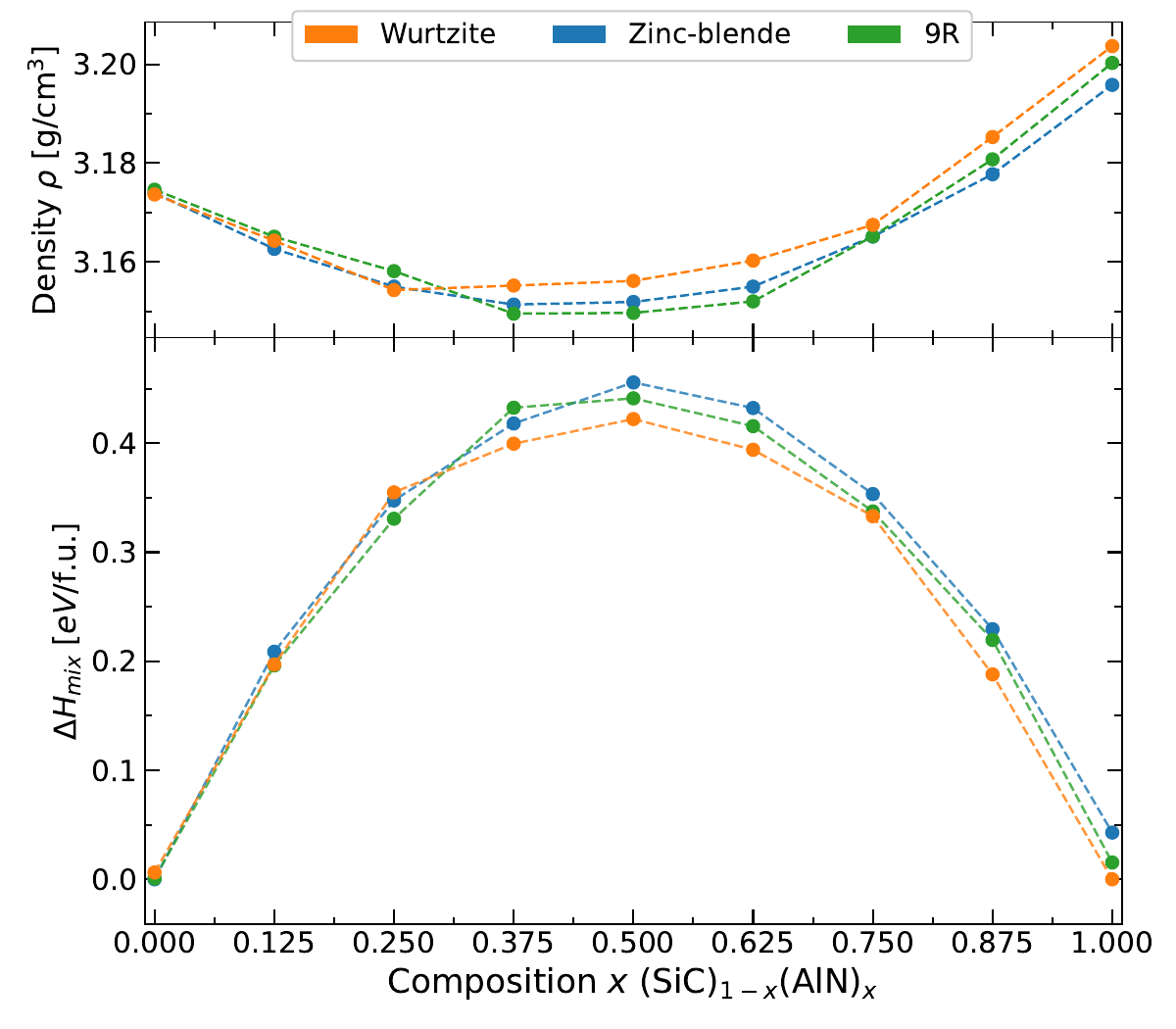}
\caption{\label{fig:hmix} Density and mixing enthalpy evolution over the entire range of the pseudo-binarry (SiC)$_{1-x}$(AlN)$_{x}$ alloy using 128 atoms SQSs in both Wurtzite (orange) and Zinc Blende (blue) phases and 144 atoms SQSs for the 9R phase.}
\end{figure}

Fig.~\ref{fig:hmix} shows density (upper) and mixing enthalpy (lower) results. In the former one can note a similar bowing of all three phases. While there is no physical reason to expect a linear evolution of the density or the lattice parameters it is always interesting to notice departure from the empirical Vegard's Law. In experiments downward bowing of the density or lattice parameters appears in some instances \cite{alloy_3,alloy_6} and doesn't in others \cite{alloy_5}, the values of the density however when compared to our results does not go beyond $2\%$ which when brought back to linear dimensions ($1\%$) is within the expected error from the PBE functional ($1-2\%$)\cite{Zhang_2018}. 

In the lower plot the mixing enthalpies are computed relative to the wurtzite and zincblende ground-states of pure AlN and SiC, respectively. Here the wurtzite phase is favored in most of the compositions, with real competition between phases only appearing for SiC rich compositions $x \leq 0.25$. This corroborates experimental results where the wurtzite phase of the (SiC)$_{1-x}$(AlN)$_{x}$ alloy is considered to be attainable at virtually any composition and is considered to dominate in most of them ($x \gtrsim 0.2$)~\cite{alloy_1,alloy_2,alloy_3,alloy_4,alloy_5,alloy_6}. These experimental results also show the importance of the growth parameters on the resulting phase of the alloy, confirming the prediciton of a relatively tight thermodynamic competition between phases.

The bowing in the density data shows that it is for the compositions with highest entropy ($x= 50\%$) that the alloy struggles the most to minimize the volume per atom which also happens to be the compositions with the highest mixing enthalpies. There is also a relation between the density and the mixing enthalpy where the denser phase appears to be energetically favored.

As indicated, the wurtzite phase prevails both in most of the mixing enthalpy results presented here but also in experimental results mentioned earlier. The wurtzite phase is also of higher interest thanks to its multiple applications mainly taking advantage of its piezoelectric properties both for AlN~\cite{AlN_1,AlN_2,AlN_3} and SiC~\cite{SiC_1,SiC_2}. Following these arguments the later part of this manuscript is focused solely on the wurtzite phase.

\subsection{Wurtzite phase\label{ssec:results_Wz}}
%
\begin{figure}
\includegraphics[width=\linewidth]{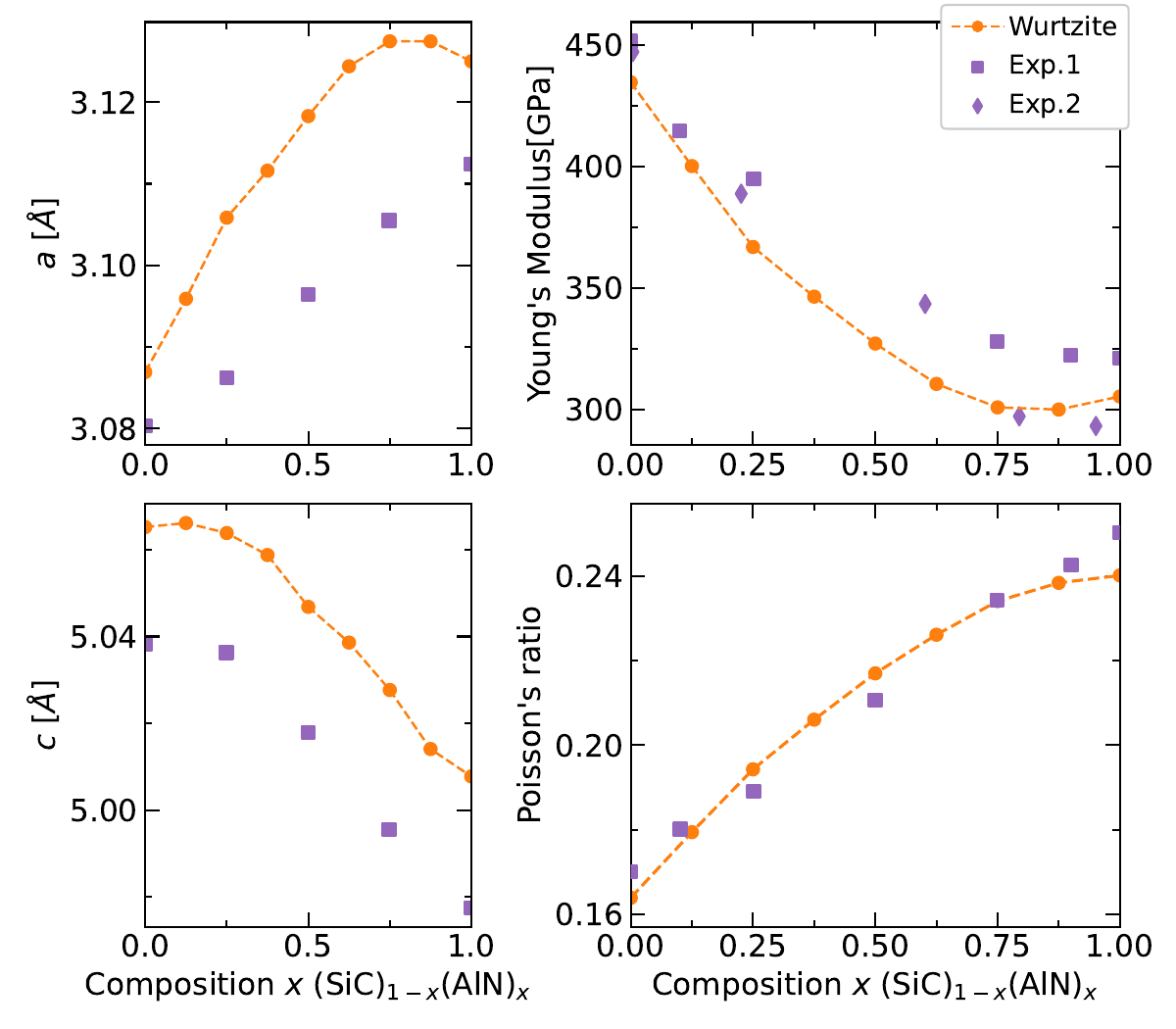}
\caption{\label{fig:lat_iso} Evolution of lattice parameters $a$ and $c$ as well as the Isotropic Young's modulus and Poisson's ratio of the Wurtzite phase of the pseudo-binary (SiC)$_{1-x}$(AlN)$_{x}$ alloy over compositions. Experimental data taken from Lubis \textit{et al.}~\cite{alloy_3} and Rafaniello \textit{et al.}~\cite{alloy_5}.}
\end{figure}

Focusing now on the wurtzite phase, we will first compare calculated properties with available experimental measurements in order to confirm the representativeness of the different SQSs, starting with lattice parameters obtained from XRD data as well as isotropic mechanical properties (Young's modulus and Poisson's ratio) obtained from the stiffness tensor. 

\begin{figure*}
\includegraphics[width=0.8\linewidth]{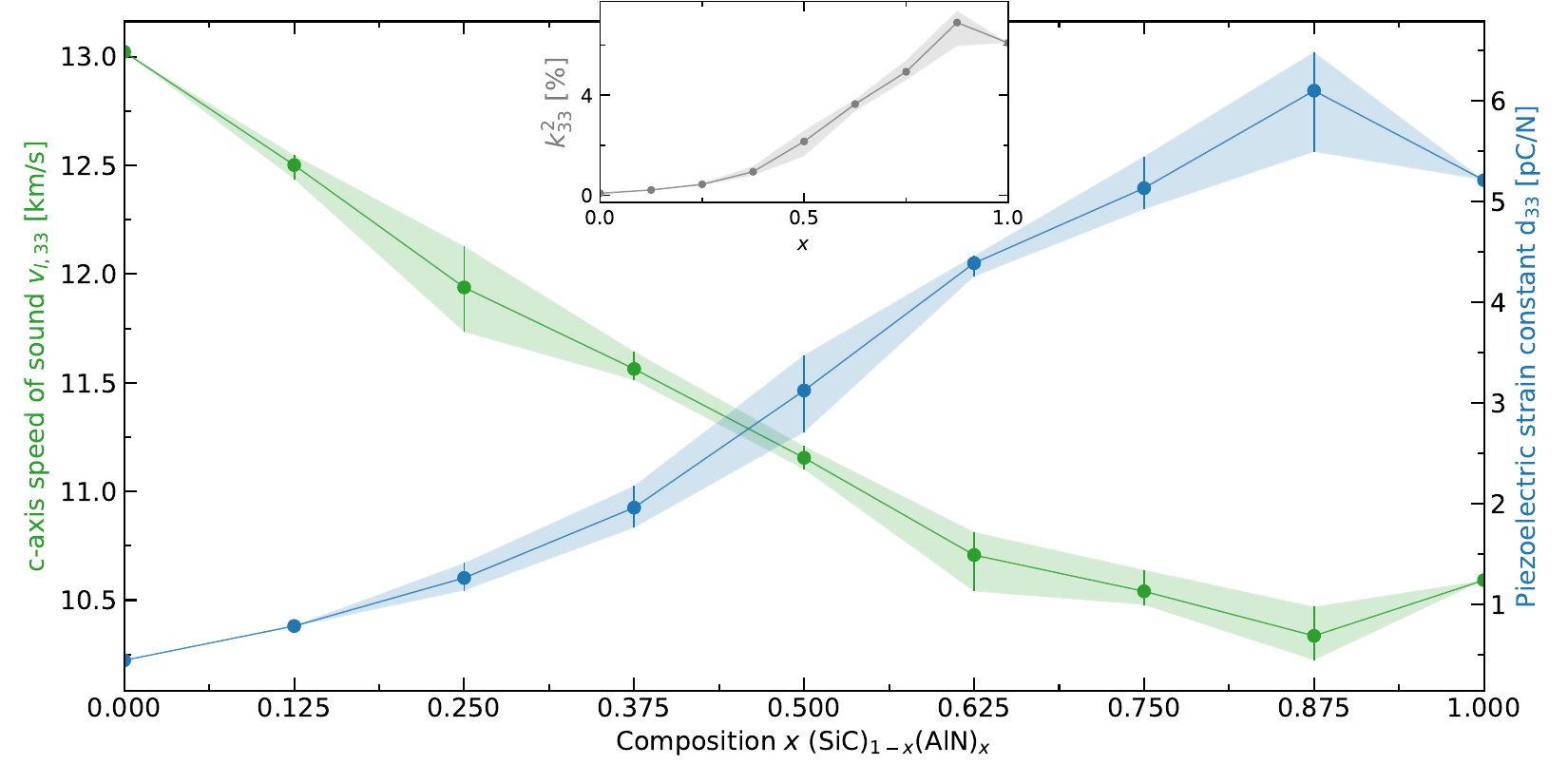}
\caption{\label{fig:mecha_props} Wurtzite pseudo-binary alloy (SiC)$_{1-x}$(AlN)$_{x}$ composition dependance of the longitudinal speed of sound along the $c$ axis (in green) and of the piezoelectric strain constant $d_{33}$ as defined in Section \ref{ssec:methods_mecha_props} (in blue). Inset shows the evolution of the electromechanical coupling coefficient $k^{2}_{33}$. Errorbars show maximal and minimal values of the sample while markers are averages. Complete stiffness, piezoelectric stress and dielectric tensors can be found in supplementary information~\cite{}}
\end{figure*}

X-ray diffracton (XRD) data are simulated for all wurtzite SQS structures, the (100) and (002) peaks are identified and their positions is used to derive the wurtzite lattice parameters similar to what is done in experiments. Comparison of the obtained values with experimental data~\cite{alloy_3,alloy_5} is presented in the left panels of Fig.~\ref{fig:lat_iso}. The lattice parameters of the SQSs show very similar behavior to experimental results throughout the different compositions. We can notably associate back the bowing of the density to the uncoupled bowing of the lattice parameters seen here, and recalculating the density from this data gives a similar curve as in Fig.~\ref{fig:hmix}. The values of the lattice parameters, when compared to the experimental ones, also showcase similar linear dimension uncertainty than the one obtained with the density ($1\%$)

The isotropic Young's modulus and Poisson's ratio are presented in the right panels of Fig.~\ref{fig:lat_iso} alongside experimental results from Refs.~\cite{alloy_3,alloy_5}. These quantities are obtained from the stiffness tensor following the procedure presented in Section~\ref{sec:methods}. Looking at the plots both Poisson's ratio and Young's modulus show very good agreement between the computational results and experiments where once again the non-linear trends are reproduced between the two. Quantitatively the Poisson's ratio data are in great agreement, where Young's modulus one show more variations with a relative difference of at most $8\%$.

Overall the agreement with experimental data shown in Fig.~\ref{fig:lat_iso} is evidence for the representativness of the theoretical methodology used here in the modeling of the pseudo-binary alloy (SiC)$_{1-x}$(AlN)$_{x}$ in the wurtzite phase and should warrant confidence in the subsequent quantities presented in this work.


As a platform for discussion we'll focus on properties relevant to the use of piezoelectric materials in Bulk Acoustic Wave (BAW) filters. In that context we present in Fig.~\ref{fig:mecha_props} results for the longitudinal speed of sound $v_{l,z}$ (\textit{green}), the piezoelectric strain constant $d_{33}$ (\textit{blue}) and the electromechanical coupling coefficient $k^{2}_{33}$, all of which oriented along the $c$ direction of the wurtzite lattice, see Section\ref{sec:methods} for definitions. The electromechanical coupling coefficient presented here is defined as~\cite{k2_def,hirata_Mat_2021}
\begin{equation}\label{eq:k2}
k^{2}_{33} \, = \, \frac{e^{2}_{33}}{\varepsilon_{33}C_{33} + e^{2}_{33}},
\end{equation}
where $\varepsilon_{33}$ is a component of the total dielectric tensor $\bm{\varepsilon}$ while $C_{33}$ and $e_{33}$ are components respectively of the stiffness and piezoelectric stress tensor. In the plot each point corresponds to an average between the calculated values of 3 different SQSs generated at that chemical composition while the error bars show maximum and minimum values of the sample.

%
\section{Discussion}\label{sec:discussion}
%
The properties presented in Fig.~\ref{fig:mecha_props} are of particular interest for applications in Bulk Acoustic Wave (BAW) filters, which initially gained interest in the early 2000s\cite{rubyEL_1999} where the acoustic resonators present in the filter were based on an AlN piezoelectric layer. Improvement of these filters was done in part through modifying the piezoelectric material and saw significant push in from the 2010s onwards with the introduction of the AlScN alloy \cite{akiyamaAM_2009} and its large increase of the piezoelectric response. 

Improvement of the piezoelectric layer can be simplified into two options, first one can improve the piezoelectric response which will improve several parameters including broadening the bandwith or increasing the efficiency of the filter, as can be seen from Eq.\ref{eq:k2}. On the other hand one can improve the speed of sound in the piezoelectric material which will lead to higher working frequencies for the filter, allow for thicker films and lower acoustic loss. 

These two approaches are usually understood as mutually exclusive as the piezoelectricity is understood as benefitting from a more ``maleable'' material where atoms can more easily move and create dipoles\cite{martinPRB_1972} while on the other hand acoustic velocities are understood to take advantage of more ``stiff'' materials with more correlated interatomic discplacements. The equations presented in Section \ref{sec:methods} can further be approximated to also reveal this inverse relationship where the speed of sound can be $v_{z}\approx\sqrt{C_{33}/\rho}$  and the piezoelectric strain response $d_{33}\approx e_{33}/C_{33}$. Here the presence of the stifness tensor coefficient $C_{33}$ on different sides of the fraction clearly shows the opposition between these two properties. 	

Curves in Fig.~\ref{fig:mecha_props} follow this understanding where $d_{33}$ and $v_{l,z}$ clearly have an inverse relationship, choice of the composition will then be dependent on the pursued type of improvment to AlN.  A higher piezoelectric response can be obtained by introducing small amounts of SiC in AlN ($x=0.875$) or on the contrary, going to $x=0.500$ one could increase the speed of sound by about $5\%$ but that would lower the piezoelectric response by about $40\%$. All the while the efficiency of the material measured by $k^{2}_{33}$ is diminishing for $x<0.875$. 

One could also try to infer the physical origins of these trends where the initial spike of the piezoelectric response when introducing SiC in AlN could be associated with the intial lowering of the density and distortion of the lattice allowing for greater independent atomic displacement especially in the c direction similarly to what can be seen in other (AlN)$_{x}$ alloys\cite{manna_PRA_2018,tasnadiPRL_2010}. However further incorporation of SiC could lower too significantly the average bond polarity and bring the piezoelectric response down towards the pure SiC meager values, one can check in fact that all bonds in the (SiC)$_{1-x}$(AlN)$_{x}$ alloy have lower polarity than the Al-N bonds.


\section{Conclusion}\label{sec:conclusion}
%
In this work we presented a first-principles study of the pseudo-binary (SiC)$_{1-x}$(AlN)$_{x}$ alloy. First, we performed First-Principles Random Structure Sampling (FPRSS) at select compositions in order to get an understanding of the relevant phases of the alloy. Wurtzite, zincblende and rhombohedral (9R) were found to be the three phases of importance as in previous bulk material results. We continued by creating large cells (128 or 144 atoms) through the use of SQSs which after DFT-PBE relaxation offered an analyses of the enthalpy of mixing competition between them. The wurtzite phase was shown to either be prevailing ($x>0.25$) or not significantly higher than the other phases. Stiffness, piezoelectric response and dielectric tensors of all the wurtzite structures were then calculated in order to obtain properties relevant to the BAW filter application. The data shows the known inverse relation between the piezoelectric strain $d_{33}$ and the acoustic velocity $v_{l,z}$ with an interesting peak of the former for $x=0.875$ and an overall downward bowing of the later, allowing for different optimization of the piezoelectric material in regards to the BAW filter application. The complete tensor data is also made available, enabling further property calculation for other MEMS applications or as parameters for lower theory level simulations.


\section{Aknowledgements}
This work was supported by the NSF PFI program under Award No. 2234617.

\bibliographystyle{apsrev4-2}
\bibliography{biblio.bib}
\end{document}